\documentclass{natureprintstyle}
\usepackage{epsfig}

\newcommand{\gtrsim}{\mathrel{\raise.3ex\hbox{$>$}\mkern-14mu
             \lower0.6ex\hbox{$\sim$}}}
\newcommand{\lesssim}{\mathrel{\raise.3ex\hbox{$<$}\mkern-14mu
             \lower0.6ex\hbox{$\sim$}}}
\newcommand{\Omegasf}{\Omega_{\mathrm{sf}}}
\newcommand{\dotOmegasf}{\dot{\Omega}_{\mathrm{sf}}}
\newcommand{\Isf}{I_{\mathrm{sf}}}
\newcommand{\fpin}{N_{\mathrm{pin}}}
\newcommand{\fmf}{N_{\mathrm{mf}}}
\newcommand{\Omegadot}{\dot{\Omega}}
\newcommand{\Omegaddot}{\ddot{\Omega}}
\newcommand{\Idot}{\dot{I}}
\newcommand{\Iddot}{\ddot{I}}
\newcommand{\Pdot}{\dot{P}}
\newcommand{\Pddot}{\ddot{P}}
\newcommand{\tauc}{\tau_{\mathrm{c}}}

\title{Rotational evolution of young pulsars due to \\ superfluid decoupling}

\author{Wynn C. G. Ho$^\ast$ \& Nils Andersson$^\ast$}

\begin{document}
\maketitle

\begin{abstract}
Pulsars are rotating neutron stars that are seen to slow down, and the
spin-down rate is thought
to be due to magnetic dipole radiation\cite{pacini68,gunnostriker69}.
This leads to a prediction for the braking index $n$, which is
a combination of the spin period and its first and second time derivatives.
However, all observed values\cite{espinozaetal11} of $n$ are below the
predicted value of $3$.
Here we provide a simple model that can explain the rotational evolution
of young pulsars, including the $n=2.51$ of the 958-year-old pulsar in
the Crab nebula\cite{lyneetal93}.
The model is based on a decrease in the effective moment of inertia due to an
increase in the fraction of the stellar core that becomes superfluid as
the star cools through neutrino emission.
The results suggest that future large radio monitoring campaigns of pulsars
will yield measurements of the neutron star mass, nuclear equation of state,
and superfluid properties.
\end{abstract}

\let\thefootnote\relax\footnote{
\textbf{{\small$\!\!\!\!\!\!\!\!\!\!$
Received 21 May 2012; accepted 15 August 2012;
published online 30 September 2012
}}}
\let\thefootnote\relax\footnote{
\begin{affiliations}
$\!\!\!\!\!\!\!\!\!\!\!\!\!$
School of Mathematics, University of Southampton, Southampton, SO17 1BJ, UK. \\
$^\ast$email: wynnho@slac.stanford.edu; n.a.andersson@soton.ac.uk
\end{affiliations}
}

The core of a neutron star has densities near and above nuclear saturation
and extends to 90\% of the radius of the star;
the remaining kilometre or so is the stellar crust.
The core is composed of degenerate matter, mostly neutrons and a small
fraction of protons and electrons (and possibly exotica, such as hyperons
and deconfined quarks, which we do not consider here).
Immediately after neutron star formation, this matter is in a normal state
due to the high temperatures reached in stellar core collapse.
However, neutron stars cool rapidly through the emission of neutrinos, and
when the temperature drops below the (density-dependent) critical temperature
for Cooper pairing, neutrons and protons form a superfluid and superconductor,
respectively\cite{migdal60,baymetal69}.
Superfluid neutrons rotate by forming quantized vortices, and the spatial
distribution of these vortices determines the rotation rate of the superfluid
core; for example, vortices migrate away from the stellar axis of rotation when
the superfluid angular velocity $\Omegasf$ decreases whereas $\Omegasf$
cannot change if the vortices are fixed in location, that is, when they are
pinned.  Meanwhile, normal matter (for example, in the crust)
rotates at an angular velocity
$\Omega$ that decreases as a result of energy loss from the stellar surface 
due to magnetic dipole radiation,
that is, $dE/dt=-\beta\Omega^4$, where $\beta\approx B^2R^6/6c^3$
and $B$ and $R$ are the neutron star magnetic field and radius,
respectively\cite{pacini68,gunnostriker69}.

A rapid decline in surface temperature was recently detected in the
330-year-old neutron star in the Cassiopeia~A supernova
remnant\cite{heinkeho10,shterninetal11}.
The observed cooling can be understood as being caused by the recent onset of
neutron superfluidity in the core of the star, combined with a much earlier
onset of proton superconductivity\cite{shterninetal11,pageetal11}.
This has provided the first direct constraints on core superfluid and
superconducting properties from neutron star observations.
These new results motivate studies of possible implications.
Here we explore the rotational evolution of young pulsars using the
newly constrained superfluid properties and assuming this superfluid core
is allowed to decouple (as discussed below).
We use simulations\cite{hoetal12} of the cooling of a neutron star
to determine the fraction of the neutron star core that is superfluid
as a function of time;
this allows us to track the normal and superfluid components of the moment
of inertia as the star ages
(see Supplementary Fig.~\ref{fig:evol}).

\begin{figure}
\centering
\includegraphics[width=7.8cm]{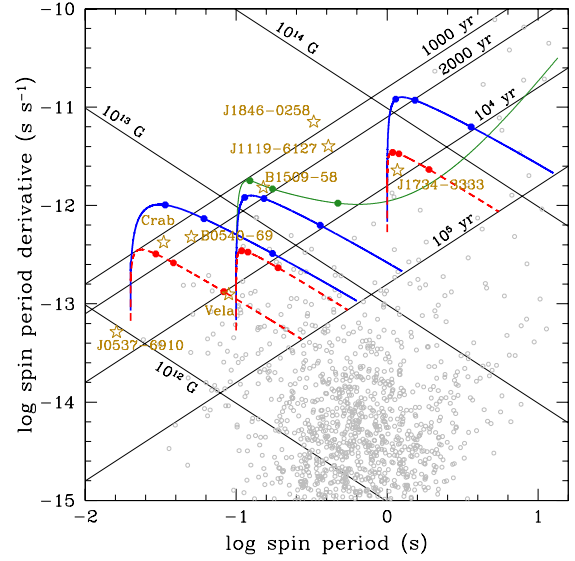}
\caption{
{\bf Pulsar spin period versus spin period derivative.}
The open circles are observed values taken from the ATNF Pulsar
Catalogue\cite{manchesteretal05}.  The
stars denote pulsars with a measured braking index (see Table~\ref{tab:psr}).
The thin diagonal lines denote characteristic age ($=P/2\Pdot$) and inferred
magnetic field [$=3.2\times10^{19}\mbox{ G }(P\Pdot)^{1/2}$].  The
curves are spin evolution tracks for pulsars with mass $1.8\,M_{\mathrm{Sun}}$
(red, dashed) and $1.4\,M_{\mathrm{Sun}}$ (blue, solid), where the spin
lag is maintained at $\Omegasf/\Omega-1\le 10^{-6}$ (by an additional
angular momentum sink);
the filled circles denote the evolution at ages 1000, 2000, and $10^4$~yr.
Also shown (green, thin solid) is the evolution of a pulsar where the spin lag
is allowed to grow.
From left to right, the initial spin period and magnetic field ($P,B$) are
taken to be
(0.02~s, $5\times 10^{12}$~G), (0.1~s, $10^{13}$~G), and (1~s, $10^{14}$~G).
These examples demonstrate that the model can explain the observed pulsar
population.
\label{fig:ppdot}}
\end{figure}

We consider a simple phenomenological model for the rotational evolution of the
normal and superfluid components of the star:
\begin{eqnarray}
\frac{d}{dt}(I\Omega) &=& - \beta\Omega^3 - \fpin - \fmf
 \label{eq:omegadot0} \\
\frac{d}{dt}(\Isf\Omegasf) &=& \fpin + \fmf, \label{eq:omegadotsf}
\end{eqnarray}
where $I$ and $\Isf$ are the moments of inertia of the normal and superfluid
components, respectively,
(and $I+\Isf=\mbox{constant}$) and $\fpin$ and $\fmf$
are torques associated with vortex pinning\cite{link03}
and dissipative mutual friction\cite{alparetal84}, respectively.
Note that it is the rotation of the normal component $\Omega$ that is
observed in pulsars.
There are three simple limits that we can consider.  The first is that
friction acts on a much shorter timescale than the spin-down
timescale\cite{alparetal84};
this is the conventional view of rotational evolution, which leads to
$\Omegasf$ closely tracking $\Omega$ and a braking index $n=3$
(where $n\equiv\Omega\Omegaddot/\Omegadot^2$ and $\dot{x}$ and $\ddot{x}$
are first and second, respectively, time derivatives of the parameter $x$),
at odds with all measured values\cite{espinozaetal11}.
The second limit is when there is no pinning or friction; we find that this
leads to $n>3$.
The final case, which we consider in detail here, is when pinning causes
$\dotOmegasf\approx 0$.  The evolution equations (\ref{eq:omegadot0}) and
(\ref{eq:omegadotsf}) can then be combined to give
\begin{equation}
\frac{d\Omega}{dt} = (\Omegasf-\Omega)\frac{1}{I}\frac{dI}{dt}
 - \beta\frac{\Omega^3}{I}. \label{eq:omegadot}
\end{equation}
The spin lag, $\Omegasf-\Omega$, is the difference in rotational velocity
between the superfluid and normal components.
Some examples of decoupled spin evolution are shown in Fig.~\ref{fig:ppdot}.

Conventional pulsar spin evolution only accounts for the second term on the
right-hand side of equation~(\ref{eq:omegadot}), with constant $B$ and $I$.
In this case, pulsars born at a particular spin period ($P=2\pi/\Omega$)
evolve by moving along lines of
constant\cite{goldreichreisenegger92,glampedakisetal11} $B$,
and the characteristic age $\tauc$ ($\equiv P/2\Pdot$)
is an estimate of the true age of the pulsar.
However, this again suggests that the conventional picture is
incomplete: in cases where an independent age can be estimated
(for example, from studying the expansion of an associated supernova remnant),
the result is often quite different from the characteristic age;
this can be seen in Table~\ref{tab:psr}.
When superfluid decoupling is taken into account, spin evolution is similar
to the conventional course, except now the moment of inertia decreases
over time.
If the spin lag remains small (for example, as a result of an angular momentum
sink acting on the superfluid),
only a small deviation (from evolution along a constant
$B$ track) is seen at intermediate and late times (after $\sim 1000-2000$~yr).
Note that we are considering intermediate times ($\sim 10^3-10^5$~yr)
in the life of a neutron star, in between the short timescale for
glitch recurrence\cite{espinozaetal11b}
($\sim 1$~yr) and the long timescales for magnetic field
diffusion\cite{goldreichreisenegger92,glampedakisetal11}
and cooling\cite{tsuruta98,yakovlevpethick04,pageetal06} ($\sim 10^5-10^6$~yr).
On the other hand, if the spin lag is allowed to become large
(for example, owing to strong vortex pinning), then
we see that a decreasing $I$ can mimic a strongly increasing magnetic field,
leading to pulsars evolving in the way suggested\cite{espinozaetal11}
for PSR~J1734$-$3333 with $n<3$.
A further departure from the conventional picture is that the characteristic
age is not an accurate indication of the true age of a pulsar at early times.

Not only are the spin period $P$ and first time derivative of the period
$\Pdot$ observable quantities, in some cases the second time derivative
$\Pddot$ can be measured.
This provides another test for our model.
Table~\ref{tab:psr} gives data on eight systems where the braking index $n$,
which is proportional to $\Pddot$, has been measured,
whereas our model predicts the braking index to be
[see equation~(\ref{eq:omegadot})]
\begin{equation}
n = 3-\frac{2\Idot}{I}\frac{\Omega}{\Omegadot}
 -\left(\frac{3\Idot}{I}\frac{\Omega}{\Omegadot}
 -\frac{\Iddot}{I}\frac{\Omega^2}{\Omegadot^2}\right)
 \left(\frac{\Omegasf}{\Omega}-1\right)
 =3-4\tauc\left|\frac{\Idot}{I}\right|, \label{eq:brake}
\end{equation}
where the second equality is obtained when $\Omegasf-\Omega\ll\Omega$.

\begin{figure}
\centering
\includegraphics[width=7.8cm]{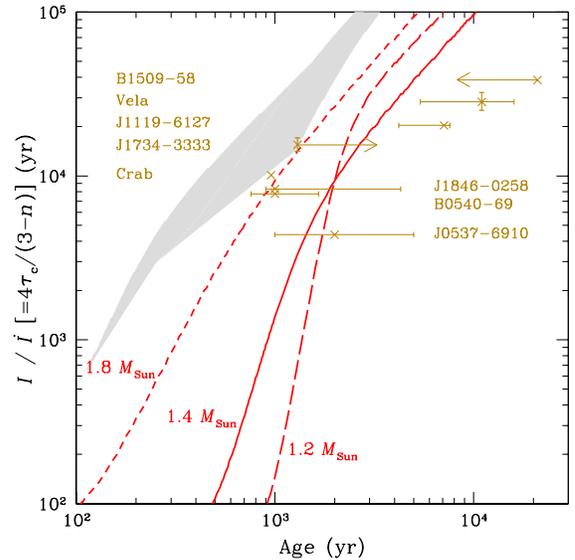}
\caption{
{\bf Constraining neutron star properties.}
The crosses denote pulsars with a measured braking index; horizontal bars and
arrows indicate uncertainties or limits on their true age, and vertical bars
are shown for pulsars with larger braking index uncertainty.
The curves show the evolution of the effective moment of inertia for pulsars
with mass $1.8\,M_{\mathrm{Sun}}$ (short-dashed),
$1.4\,M_{\mathrm{Sun}}$ (solid), and $1.2\,M_{\mathrm{Sun}}$ (long-dashed),
assuming a particular stellar and superfluid model\cite{pageetal11},
and the shaded region represents an alternative model\cite{shterninetal11};
both sets of superfluid parameters fit the rapid cooling seen in the
Cassiopeia~A neutron star (see Supplementary Information)
and provide examples of the effect of different $|I/\Idot|$ models
[see equation~(\ref{eq:brake})].
These moment of inertia evolution curves are analogous to the thermal cooling
curves\cite{tsuruta98,yakovlevpethick04,pageetal06} used to determine
properties of neutron stars.
\label{fig:timescales}}
\end{figure}

As $\tauc$ and $n$ are observable quantities
(related simply to $P$, $\Pdot$, and $\Pddot$) for a given pulsar,
we can compare our predictions to the pulsars in Table~\ref{tab:psr};
this is shown in Fig.~\ref{fig:timescales}.
For the Crab pulsar (the only one with a known age), we infer a relatively
high mass of $\approx 1.8\,M_{\mathrm{Sun}}$ (for the particular equation
of state and superfluid properties we consider; see Supplementary Information).
Furthermore, we can use the mass determination to estimate the initial period
and magnetic field of the pulsar,
and we find an initial period $\sim 0.02$~s and $B\sim4\times 10^{12}$~G.

The fact that our simple model is able to explain the observed pulsar
properties demonstrates the merits of the notion of decoupled spin-evolution
due to the onset of core superfluidity. However,  key questions remain to be
understood.  The main assumption in our model is that core superfluid
neutrons are allowed to decouple and pin. Core pinning is thought to be the
result of the interaction between superfluid vortices and fluxtubes in the
proton superconductor\cite{link03}. Whether this mechanism can be strong
enough to act in the way assumed in our analysis is not clear at this time.
Theoretical work is also required to determine whether the spin-lag
$\Omegasf-\Omega$ can be kept small during the evolution by an (at this time)
unspecified angular momentum sink.
From an observational point-of-view, discovery and long-term
monitoring of a large number of systems by radio telescopes such as LOFAR
and SKA will allow accurate timing of many young pulsars in the future.
Taking these pulsars as an ensemble, we can constrain the nuclear equation of
state and superfluid properties (because these determine the evolution of the
moment of inertia),
analogous to what is done in studies of neutron star thermal evolution.
Knowledge of these properties can then be used to infer the mass of individual
pulsars.
Finally, radio timing measurements may be able to constrain neutron star
thermal evolution, independently of measurements at X-ray energies.

\medskip
\noindent {\it Note added in proof.}
After completion of this work, we became aware of the independent development
of a similar mechanism for variation of the pulsar-braking index that was
presented by E.~Kantor at two conferences in 2011 (http://go.nature.com/sxaWeN;
http://go.nature.com/veAOt1).
\\

\begin{table*}
\small{
\begin{tabular}{lcccccc}
\hline
Pulsar & Supernova & Period & Period derivative & Characteristic & Braking
 & Age \\
name & remnant & (s) & (s s$^{-1}$) & age $\tauc$ (yr) & index $n$ & (yr) \\
\hline
B0531$+$21 & Crab & 0.0331 & 4.23$\times 10^{-13}$ & 1240 & 2.51(1) (ref. 4) & 958 \\
J0537$-$6910 & N157B & 0.0161 & 5.18$\times 10^{-14}$ & 4930 & $-1.5$(1) (ref. 20) & 2000$^{+3000}_{-1000}$ (ref. 21) \\
B0540$-$69 & 0540$-$69.3 & 0.0505 & 4.79$\times 10^{-13}$ & 1670 & 2.140(9) (ref. 22) & 1000$^{+660}_{-240}$ (ref. 23) \\
B0833$-$45 & Vela & 0.0893 & 1.25$\times 10^{-13}$ & 11300 & 1.4(2) (ref. 24) & 11000$^{+5000}_{-5600}$ (ref. 25) \\
J1119$-$6127 & G292.2$-$0.5 & 0.408 & 4.02$\times 10^{-12}$ & 1610 & 2.684(2) (ref. 26) & 7100$^{+500}_{-2900}$ (ref. 27) \\
B1509$-$58 & G320.4$-$1.2 & 0.151 & 1.54$\times 10^{-12}$ & 1550 & 2.839(3) (ref. 22) & $<21000$ (ref. 28) \\
J1846$-$0258 & Kesteven 75 & 0.325 & 7.08$\times 10^{-12}$ & 729 & 2.65(1) (ref. 22) & 1000$^{+3300}_{-100}$ (ref. 29) \\
J1734$-$3333 & G354.8$-$0.8 & 1.17 & 2.28$\times 10^{-12}$ & 8120 & 0.9(2) (ref. 3) & $>1300$ \\
\hline
\end{tabular}
\caption{
{\bf Pulsars with observed braking index.}
The periods and period derivatives are taken from ref.~19.
The numbers in parentheses show braking index uncertainty in the last digit.
For J1734$-$3333, we give a lower limit of the age, which we estimate by
considering the supernova remnant size
(21~parsecs; ref.~30)
and remnant expansion velocity $v_{\mathrm{SNR}}$, to obtain an age
$\sim 2000\mbox{ yr }(10^4\mbox{ km s$^{-1}$}/v_{\mathrm{SNR}})$;
and considering the pulsar's distance away from the centre of the
supernova remnant (46~parsecs; ref.~30)
and pulsar space velocity $v_{\mathrm{pulsar}}$,
to obtain an age
$\sim 23000\mbox{ yr }(2000\mbox{ km s$^{-1}$}/v_{\mathrm{pulsar}})$.}
\label{tab:psr}
}
\end{table*}

\smallskip
\noindent {\small {\bf Acknowledgements} \\
W.C.G.H. thanks D. Yakovlev for providing equation of state tables.
W.C.G.H. appreciates the use of the computer facilities at the Kavli
Institute for Particle Astrophysics and Cosmology.
W.C.G.H. and N.A. acknowledge support from the Science and Technology
Facilities Council (STFC) in the United Kingdom.
}

\smallskip
\noindent {\small {\bf Author contributions} \\
W.C.G.H. contributed to developing the model, performed the calculations,
and wrote the manuscript.
N.A. contributed to developing the model and writing the manuscript.
}

\smallskip
\noindent {\small {\bf Additional information} \\
Supplementary information is available in the online version of the paper.
Reprints and permissions information is available online at
www.nature.com/reprints.
Correspondence and requests for materials should be addressed to W.C.G.H.
or N.A.
}

\smallskip
\noindent {\small {\bf Competing financial interests} \\
 The authors declare no competing financial interests.
}

\clearpage

\pagestyle{empty}
\renewcommand{\thesubsection}{\Alph{subsection}}
\renewcommand{\thefigure}{\thesubsection\arabic{figure}}
\renewcommand{\thetable}{\thesubsection\arabic{table}}
\setcounter{subsection}{19}
\setcounter{figure}{0}
\setcounter{table}{0}
\section*{Supplementary Information}
\subsubsection*{Neutron star cooling and moment of inertia evolution}

Neutron stars begin their lives very hot (with temperatures
$T>10^{11}\mbox{ K}$) but cool rapidly through the emission of neutrinos.
Neutrino emission processes depend on uncertain physics at the supra-nuclear
densities ($\rho>2.8\times 10^{14}\mbox{ g cm$^{-3}$}$) of the neutron star
core\cite{tsuruta98,yakovlevpethick04,pageetal06}.
The recent observation of rapid cooling\cite{heinkeho10,shterninetal11}
of the neutron star in the Cassiopeia~A supernova remnant provides the first
constraints on the (density-dependent) critical temperatures for the onset of
superfluidity of core protons $T_{\mathrm{cp}}$ (in the singlet state) and
neutrons $T_{\mathrm{cnt}}$ (in the triplet state), that is,
$T_{\mathrm{cp}}\sim (2-3)\times 10^9$~K and a maximum
$T_{\mathrm{cnt}}\approx 5\times 10^8$~K
for a superfluid neutron model at relatively shallow densities\cite{pageetal11}
or a maximum $T_{\mathrm{cnt}}\approx(7-9)\times 10^8$~K
for a superfluid neutron model at deep densities\cite{shterninetal11}.

\begin{figure}[htb]
\centering
\includegraphics[width=8.0cm]{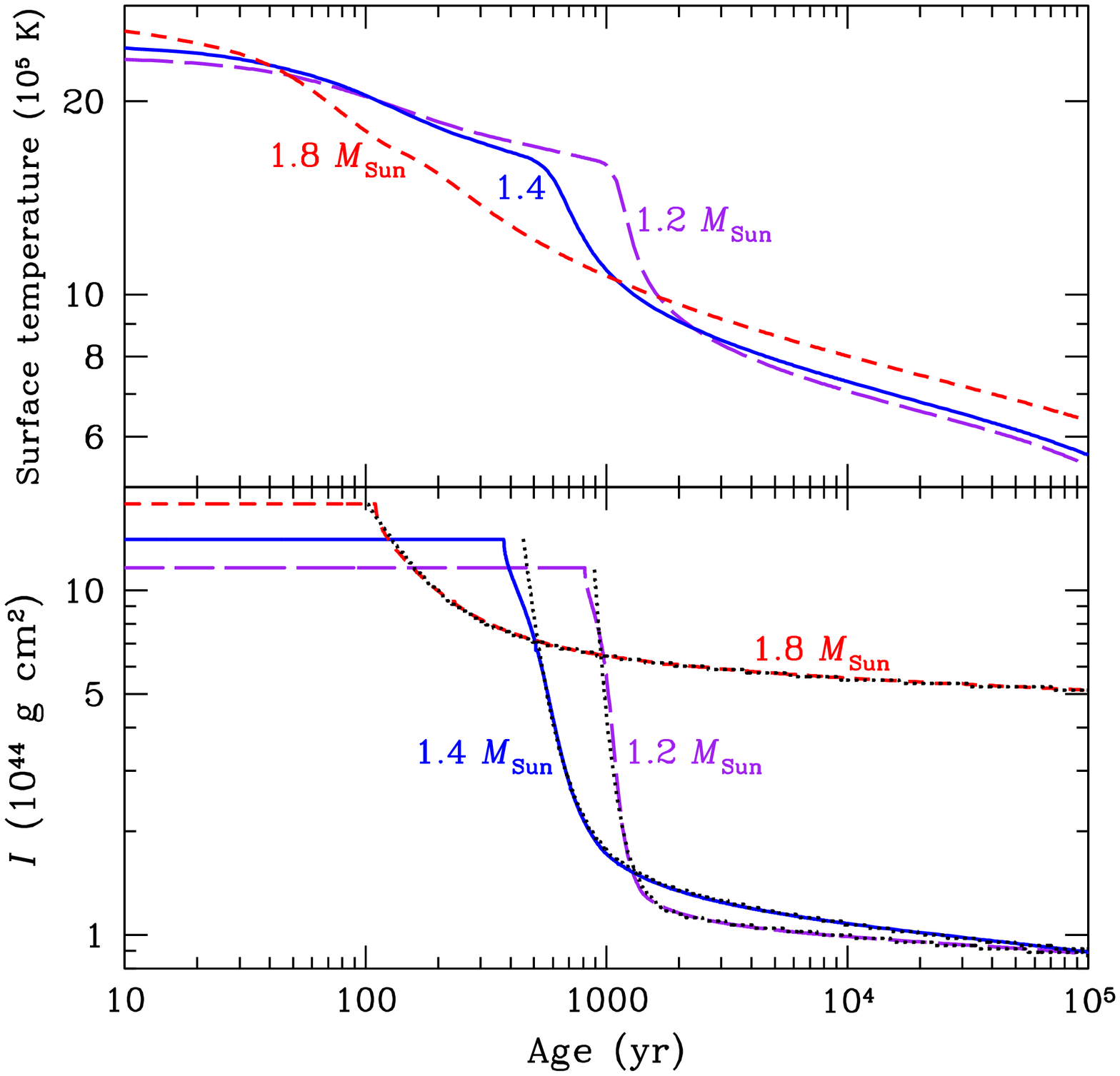}
\caption{
{\bf Thermal and moment of inertia evolution of neutron stars.}
The curves show the evolution of the surface temperature (top) and moment of
inertia (bottom) for neutron stars with mass $1.8\,M_{\mathrm{Sun}}$
(red, short-dashed), $1.4\,M_{\mathrm{Sun}}$ (blue, solid), and
$1.2\,M_{\mathrm{Sun}}$ (purple, long-dashed).
The dotted curves show analytic fits using equation~(\ref{eq:ifit}).
\label{fig:evol}}
\end{figure}

We calculate the thermal evolution of neutron stars by solving the
relativistic equations of energy balance and heat flux.
We use a stellar model based on the APR equation of state and consider
either the shallow or deep model for triplet neutron pairing in the
core\cite{hoetal12}.
Note that we improve upon the calculations of ref.~10 by using
more accurate ion and electron heat capacities\cite{potekhinchabrier10}.
Due to high thermal conductivity, neutron stars after $\sim 10-100$~yr
are essentially isothermal\cite{lattimeretal94,gnedinetal01,yakovlevetal11};
the exact time is unimportant for our work since the youngest neutron star
we consider has an age $\approx 1000$~yr (see Table~\ref{tab:psr}).
The evolution of the (gravitationally redshifted) surface temperature
for models with the shallow neutron superfluid and an unmagnetized iron
envelope is shown in Fig.~\ref{fig:evol}.
Enhanced neutrino emission due to Cooper pairing of neutrons begins to occur
at the onset of superfluidity in the core.
At an age of a few hundred to one thousand years, large portions of the
(predominantly neutron) stellar core become superfluid, and rapid cooling
occurs.

We calculate the moment of inertia of the normal and superfluid components
by\cite{ravenhallpethick94}
\begin{equation}
I = \frac{8\pi}{3}\int_{0}^{R}(\rho+P/c^2)\Lambda r^4 dr,
\end{equation}
where $P$ is the local pressure, $\Lambda=(1-2Gm/c^2r)^{-1}$, and $m$ is the
mass enclosed within $r$.
Note that we neglect for simplicity a term that approximately accounts
for relativistic frame-dragging\cite{ravenhallpethick94}, which would
only change $I$ by $<10\%$.
Fig.~\ref{fig:evol} shows the evolution of the normal component of $I$.

In order to facilitate numerical calculations of $\Idot(t)$ and $\Iddot(t)$,
we fit $I(t)$ with the following function
\begin{equation}
\log I = a_3\tanh[a_2(\log t-a_1)]+(\log t-b_1)^{b_2}+c_1, \label{eq:ifit}
\end{equation}
where the fit parameters $a_1, a_2, a_3, b_1, b_2, c_1$ are given in
Table~\ref{tab:fitparam}.
A comparison of the approximate fit to the exact result is shown in
Fig.~\ref{fig:evol}.  Note that the largest deviations occur briefly at
(early) times that are not of current relevance.

We note that we only consider a single model for the nuclear equation of
state and two models for the neutron superfluid gap.
Exploration of a larger range of models
(in order to obtain, for example, error bars on neutron star mass)
is required and the subject of future work.
\\

\begin{table}[htb]
\small{
\begin{tabular}{ccccccc}
\hline
Neutron star mass & $a_1$ & $a_2$ & $a_3$ & $b_1$ & $b_2$ & $c_1$ \\
\hline
1.2 $M_{\mathrm{Sun}}$ & 2.90 & 7.4 & -1.5 & 0.10 & -1.0 & 45.24 \\
1.4 $M_{\mathrm{Sun}}$ & 2.59 & 4.6 & -1.2 & 0.95 & -1.2 & 44.97 \\
1.8 $M_{\mathrm{Sun}}$ & 1.98 & 2.6 & -0.31 & 0 & -1.6 & 44.945 \\
\hline
\end{tabular}
\caption{
{\bf Fit parameters for moment of inertia evolution.}
}
\label{tab:fitparam}
}
\end{table}

\end{document}